# Synergizing superresolution optical fluctuation imaging with single molecule localization microscopy

Shachar Schidorsky[1,3], Xiyu Yi[2,3], Yair Razvag[1], Yonatan Golan[1], Shimon Weiss[2]*, Eilon Sherman[1]*

[1] *Racah Institute of Physics, The Hebrew University, Jerusalem, Israel, 91904*

[2] *Department of Chemistry & Biochemistry, UCLA, CA, USA, 90095-1569*

[3] *Equal contribution*

*\* Corresponding authors*

**Single molecule localization microscopy (SMLM) techniques enable imaging biological samples well beyond the diffraction limit of light, but they vary significantly in their spatial and temporal resolutions. High-order statistical analysis of temporal fluctuations as in superresolution optical fluctuation imaging (SOFI) also enable imaging beyond diffraction limit, but usually at a lower resolution as compared to SMLM. Since the same data format is acquired for both methods, their algorithms can be applied to the same data set, and thus may be combined synergistically to improve overall imaging performance. Here, we find that SOFI converges much faster than SMLM, provides additive information to SMLM, and can efficiently reject background. We then show how SOFI-assisted SMLM imaging can improve SMLM image reconstruction by rejecting common sources of background, especially under low signal-to-background conditions. The performance of our approach was evaluated using a realistic simulation of fluorescence imaging we developed and further demonstrated on experimental SMLM images of the plasma membrane of activated fixed and live T cells. Our approach significantly enhances SMLM performance under demanding imaging conditions and could set an example for synergizing additional imaging techniques.**

**Introduction**

Multiple super-resolution microscopy (SRM) techniques currently enable imaging biological samples well beyond the diffraction limit of light[1]. Such techniques may vary significantly not only in their spatial and temporal resolutions, but also in their performance with regard to background rejection, sample photobleaching, and potential image reconstruction artefacts. For instance, single molecule localization microscopy (SMLM) techniques such as Photoactivated Localization Microscopy (PALM)[2] and (direct) Stochastic Optical Reconstruction Microscopy (dSTORM)[3,4] can localize single molecules with resolution down to ~10-20nm, but require thousands of frames for the collection of data for a single image. In contrast, methods such as Stimulated Emission Depletion (STED)[5], Structured Illumination Microscopy (SIM)[6] and nonlinear SIM[7] perform measurements on ensembles of fluorophores, require scanning, and provide spatial resolution of ~10-100nm[1] in biological samples.

The variability in the characteristics and performance of such techniques that can be applied to the same experimental data set may allow for the synergetic combination of these techniques. Such combination can potentially yield faster acquisition and analysis, more optimized extraction of information from experimental data, and new insights into biophysical processes under study. Specifically, SMLM and super-resolution optical fluctuation imaging (SOFI)[8] share a wide-field imaging configuration and use the reversible blinking of fluorophores for generating a super-resolved image. Thus, these techniques can be applied to the same data set. However, different fluorophore densities and signal-to-background ratios (SBRs) are required for their respective optimal performances.

Nonetheless, we argue, and demonstrate below, that these methods could be combined synergistically to improve overall imaging performance. Specifically we focus on the combination of the techniques of PALM, dSTORM and SOFI[8]. In our study, we found that image reconstruction by SOFI converges much faster than by SMLM, that it provides uncorrelated information to SMLM, and that it can efficiently reject background in SMLM imaging. We then demonstrate how SOFI-assisted SMLM imaging can improve SMLM image reconstruction by rejecting common sources of background.

**Results**

Image acquisition requires time for the collection of information. Super resolution imaging increases the extent of acquired information in the image and thus, takes longer to acquire than diffraction limited imaging techniques. Fig. 1A shows the time-dependent image build-up using either 2nd order SOFI (top row) or PALM (bottom row) image reconstruction of a fixed T-cell expressing Dronpa-actin and spread on an activating αCD3-coated coverslip. The pronounced actin ring at the cell interface with the coverslip is a hallmark of T cell activation and becomes quickly visible via SOFI image reconstruction, but takes longer to show via PALM image reconstruction. Fig. 1B shows the time-dependent image build-up using either 2nd order SOFI (top row) or dSTORM (bottom row) image reconstruction of a fixed T cell stained with a primary antibody against the phosphorylated and active form of the kinase ZAP70 (αpZAP70-Alexa647). ZAP70 gets recruited to the T cell antigen receptor (TCR) upon TCR engagement with a cognate antigen, where it gets phosphorylated and activated, and further propagates the TCR signalling cascade by phosphorylating downstream signalling proteins [9]. Importantly, ZAP70 shows pronounced sub-diffraction clusters at the PM that form as a result of T cell activation [10]. Again, SOFI images seem to converge faster to the final image than dSTORM images. Thus, SOFI image reconstruction seem to converge faster than SMLM imaging [11].

An effective way for evaluating the differences in performance between two imaging techniques is to pixel-wise compare parameters of the reconstructed images. Particularly, the statistical dependence of SMLM parameters (characteristic of peak centroiding) on 2nd order SOFI pixel value (in which SMLM peaks are found) can be assessed by plotting 2D scatter plots of SMLM parameters vs. SOFI pixel values (as in Fig. 1A). Such parameters included the localization statistics of SMLM peaks (variance σ and localization uncertainty $\hat{\sigma}_{x,y}$), their intensity, and their local background statistics. For analyses, we either used the PALM data of T cells expressing Dronpa-actin presented in Fig. 1A or dSTORM images of T cells stained for pZAP70. The SOFI value of a given pixel reports on the extent of temporal correlation of intensity fluctuations within this pixel. Note that each SOFI pixel value represents a correlation amplitude that could originate from a superposition of fluctuating signals from multiple emitters. The same emitters would result in multiple SMLM localizations within that pixel or in its vicinity, but on a much finer grid. The 2D scatter plots of Fig. 2 show that SOFI values do not correlate well with any of the SMLM localization parameters (for PALM see Fig. 2A,B and for dSTORM see Fig. 2C,D). The low correlation is attributed to the difference in pixel-wise analysis for SOFI vs. centroid-based analysis for SMLM. Thus, although both methods yield similar images of the underlying morphologies, SOFI provides uncorrelated and additional information to the information extracted from SMLM. Moreover, we find that the added SOFI information allows for efficient isolation and classification of different emitter types in the sample, and most importantly, the efficient separation of background from signal. For example, the green dots in Fig. 2A highlight the cell footprint, where Dronpa-actin molecules can be identified, while blue dots mark false localizations from the background, outside of the cell footprint. Red dots in Fig. 2A and B correspond to a subpopulation of localizations, having high $\sigma$ and high intensity and likely represent aggregates or miss-localizations due to high density of emitters. We further discuss this point below.

Experimental data often suffers from significant background (autofluorescence), out-of-focus background, varying concentrations of fluorophores within the same field of view (FOV), complicated and heterogeneous photophysics of fluorophores, cellular dynamics (when imaging live samples), and added noise from the detectors. These factors often complicate the localization of single fluorophores, as in the case of PALM and dSTORM. As a result, optimization of localization parameters is often required in order to maximize the probability of detection of single emitters ($P_d \stackrel{\text{def}}{=} \frac{\text{detected flourophores}}{\text{total \# flourophores}}$) and minimize the probability of false detection ($P_{fd} \stackrel{\text{def}}{=} \frac{\text{false detections}}{\text{total \# detections}}$). We argue that the temporal correlation of a single emitter signal (encoded as a SOFI pixel value) can assist in accepting or rejecting a peak localization event (Fig. 2). To test this approach, we developed a realistic simulator and critically tested SOFI-assisted background rejection on simulated data (as described below and in a dedicated section in the SI).

In SMLM imaging, it is often hard to distinguish the signal from background as the positions of emitters are not a-priori known. Simulated data provided ground-truth information that background rejection algorithms can be tested against. We therefore developed realistic simulations of fluorescence microscopy imaging data. The simulations encoded realistic point spread function (PSF) based on Gibson & Lanni's model [12]. The simulation also modelled TIRF or epi illumination, as well as shot noise and a realistic pixel size to model detection optics and camera. The simulator also allows for the inclusion of different emitter types (emission wavelength and photophysics parameters), different density and binding uncertainties of emitters to feature of interest, variable sample background, out-of-focus light, sample auto-fluorescence and bleaching dynamics. All parameters could be included and adjusted to match real data sets and imaging conditions (see Fig. S1 and SI for further details). The simulator provides a powerful tool to generate a dataset library covering various experimental conditions. In turn, background rejection algorithms can be tested against large parameter space and compared to the ground-truth, allowing for their optimization and refinement.

The image reconstruction of SMLM images has so far employed subjective thresholding to identify single emitters and to reject surrounding background. The simulator allowed us to develop an objective and automated thresholding algorithm. As a representative dataset, we simulated single emitters that highlighted a 3D (or 2D) tubulin-like mesh and included realistic background components (see Fig. 3B, S1A,B and SI for further details on simulation parameters). We then analysed the dataset using SMLM and SOFI and plotted histograms of SOFI (Fig. 3A) or sum-intensity (Fig. S1A) values of all pixels in the images. A distinct sub-population of background pixels (lowest values) could be identified in the log-normal histogram (Fig. 3A). This grouping of pixels into a distinct background subpopulation seemed robust to various simulated parameters, including imaging in 3D and 2D (Fig. S1), and could not be achieved by using only the sum image (compare Fig. 3A and S2A). Importantly, this grouping enables us to automatically detect the background peak in the histogram and fit it with a Gaussian. Next, a threshold could be placed based on SOFI values (see further details in the section on data analyse in the SI). This threshold formed a spatial mask that efficiently isolated the fraction of ground-truth emitters (i.e. the true signal) in the SMLM image, while allowing a predetermined percentage of (background-originated) false-positive detections, i.e. false SMLM localizations, to 'leak' through the mask (compare panels B and C in Fig. 3). Similar masking based on a sum-image threshold (Fig. S2B,C) resulted in a wider and less efficient background rejection as compared to SOFI-based masking. To quantify the performance of background rejection using either approach, we studied how the true and false detection rates were affected by the threshold value. Lowering the SOFI threshold increased the $P_{fa}$ while increasing the detection efficiency, i.e. decreasing the undetected fraction of ground-truth emitters (Fig. 3D, blue line). In contrast, masking the SMLM image by setting a threshold on the sum-image achieved inferior $P_d$ vs. $P_{fa}$ curves in comparison to the SOFI-based approach (Fig. 3D, red line). The sum-image mask could achieve high $P_d$ only at much higher $P_{fa}$ values. Moreover, the $P_d$ vs. $P_{fa}$ curve for the sum image masking is not monotonous since the sum image mask confuses signal pixels with background pixels (e.g. compare Fig. S2B and C).

Importantly, the SOFI-based spatial mask could be used to significantly accelerate SMLM image reconstruction, since the laborious task of peak identification and fitting could now be focused on a small part of the imaging field. For instance, we found that SOFI based filtering of simulated SMLM data (Fig.3D) that was set for 0.8% false positive detections ($P_{fa}$) could achieve 93.3% probability of true detections ($P_d$) with a mask size of only 2.6% of the imaged field.

We next applied our SOFI-assisted SMLM image reconstruction to experimental data, acquired by PALM (Fig. 4A-4C) and dSTORM (Fig. 4D-4F). In both cases we found that automatic SOFI-based thresholding and detection showed efficient background rejection that could not be obtained by sum-image based thresholding (Fig. S2D-J, and as shown earlier for the simulated data in Fig. S2A-C).

So far, we have presented results for fixed cells. In order to employ our technique of SOFI-assisted background rejection to SMLM of live cells, we first needed to evaluate the speed by which the SOFI mask converges. For that, we first calculated the probability of detection ($P_d$) of a SOFI mask that was dynamically updated during image acquisition of simulated 3D data (Fig. 5A; simulated data shown in Fig. 3B) for a constant (5%) background fraction. The SOFI mask was calculated for a sliding window of 100 frames with time steps of 40 frames (as in Fig. 3A; for a detailed discussion of the background fraction see SI section on data analyses). We found that the SOFI mask converged quickly relative to the movie length, jumping from a $P_d$ value of ~0.55 to ~0.87 within 80 frames, and saturating at ~0.93 after 200 frames. Thus, we conclude that a SOFI mask over 100 frames around each time point could provide an efficient way of background rejection in live cell SMLM. To demonstrate the effectiveness of our approach, we employed a similar mask (calculated every 100 frames) to PALM images of a representative live T-cell expressing TCRζ-Dronpa and spread on an αCD11a-coated coverslip. The sequence of original and SOFI-filtered PALM images are shown in Fig. 5B and supplemental movie M1, and demonstrate efficient background rejection by the dynamic SOFI masking, even when significant spatio-temporal dynamics appears in the data. The fastest dynamics that can be achieved by our technique is limited by the need to construct the SMLM image (~220 frames in Fig. 5B), rather than the relatively faster SOFI mask calculation (of 100 frames). Notably, we found that out-of-focus background often broadened the cell features in live cell SMLM imaging, while SOFI-assisted rejection of such background significantly narrowed the size of these features (compare insets of images in top vs. middle row in Fig. 5B and see representative intensity profiles in the bottom row).

**Discussion**

SMLM and SOFI share a wide-field imaging configuration and use the reversible blinking of fluorophores for generating a super-resolved image. By performing SOFI and SMLM image reconstruction on the same data-sets and comparing SOFI values and SMLM localization parameters, we identified the relatively fast convergence of SOFI relative to SMLM, its uncorrelatedness to SMLM, and its effective background rejection capability. These properties offer the opportunity to combine SMLM with SOFI image reconstruction for optimized SMLM performance. Benefits of SOFI-assisted SMLM image reconstruction and analytic tools include: efficient background rejection with automatic thresholding, the isolation of emitter subpopulations, and the acceleration of SMLM image reconstruction. To demonstrate the advantages of our approach, we applied it to both PALM and dSTORM experimental data. SOFI and dSTORM have been previously compared under various fluorophores blinking statistics[11] and SMLM was compared under different labelling density conditions [13]. In our study, we also developed a realistic simulator for wide-field fluorescence imaging to critically evaluate the performance of our approach and to develop automatic SOFI-based thresholding for SMLM. Our algorithms and simulations are available on line (see link in the description of Supporting Information below).

Both high-order SOFI and SMLM image reconstructions are computationally expensive. However, the computationally inexpensive 2nd order SOFI is already sufficient for efficient background rejection. With further improvements in computer hardware, the development and implementation of more optimal and cost-effective SMLM and SOFI algorithms and their possible implementation in graphical processing units (GPUs) will facilitate faster and possibly even real-time image reconstruction of the SOFI-assisted SMLM technique.

The uncorrelatedness between SOFI and SMLM points to a more general aspect of image acquisition. It demonstrates that every imaging technique is only an estimated version of the true data and carries with it its own set of characteristics artefacts. For instance, the detection of SMLM peaks and their fitting with Gaussians introduce errors and artefacts. SOFI image reconstruction requires a different set of information (intensity fluctuations) and considerations. Thus, the artefacts generated by either technique can be compensated for (and possibly corrected) by tracing back the loss or distortion of information by each technique. Thus, our approach could set an example for the synergetic combination of additional SRM and diffraction-limited imaging techniques.

To conclude, the combination of the techniques described herein could not only improve significantly SMLM image reconstruction, but also sets common tools for the evaluation, comparison and synergistic combination of various imaging techniques.

**Supporting Information Available:** Details on Materials and Methods include information on cell cloning, and cell line handling, sample preparation for PALM and dSTORM, PALM and dSTROM microscopy, SOFI analyses and SOFI-assisted SMLM analyses. We also devote a separate section for the realistic simulation of fluorescence microscopy. Supplemental Figures S1-S2 depicting (S1) Realistic simulations of fluorescence imaging, and (S2) Inefficient filtering of SMLM detections using sum-image thresholding. Supplemental movie M1 showing unfiltered vs SOFI-assisted SMLM imaging of a live T cell, expressing TCRζ-Dronpa spreading on a coverslip. The code of the developed algorithms and sample data are available online (https://github.com/ShermanLab/enhanced-widefield-sr).

**References**

1. Hell, S. W. Far-field optical nanoscopy. *Science* **316**, 1153-1158, (2007).
2. Betzig, E. *et al.* Imaging intracellular fluorescent proteins at nanometer resolution. *Science* **313**, 1642-1645, (2006).
3. Rust, M. J., Bates, M. & Zhuang, X. Sub-diffraction-limit imaging by stochastic optical reconstruction microscopy (STORM). *Nat Methods* **3**, 793-795, (2006).
4. van de Linde, S. *et al.* Direct stochastic optical reconstruction microscopy with standard fluorescent probes. *Nat Protoc* **6**, 991-1009, (2011).
5. Hell, S. W. & Wichmann, J. Breaking the diffraction resolution limit by stimulated emission: stimulated-emission-depletion fluorescence microscopy. *Opt Lett* **19**, 780-782, (1994).
6. Gustafsson, M. G. L. Surpassing the lateral resolution limit by a factor of two using structured illumination microscopy. *Journal of Microscopy-Oxford* **198**, 82-87, (2000).
7. Gustafsson, M. G. Nonlinear structured-illumination microscopy: wide-field fluorescence imaging with theoretically unlimited resolution. *Proceedings of the National Academy of Sciences of the United States of America* **102**, 13081-13086, (2005).
8. Dertinger, T., Colyer, R., Iyer, G., Weiss, S. & Enderlein, J. Fast, background-free, 3D super-resolution optical fluctuation imaging (SOFI). *Proc Natl Acad Sci U S A* **106**, 22287-22292, (2009).
9. Wang, H. *et al.* ZAP-70: an essential kinase in T-cell signaling. *Cold Spring Harb Perspect Biol* **2**, a002279, (2010).
10. Sherman, E. *et al.* Functional nanoscale organization of signaling molecules downstream of the T cell antigen receptor. *Immunity* **35**, 705-720, (2011).
11. Geissbuehler, S., Dellagiacoma, C. & Lasser, T. Comparison between SOFI and STORM. *Biomed Opt Express* **2**, 408-420, (2011).
12. Gibson, S. F. & Lanni, F. Experimental Test of an Analytical Model of Aberration in an Oil-Immersion Objective Lens Used in 3-Dimensional Light-Microscopy. *Journal of the Optical Society of America a-Optics Image Science and Vision* **8**, 1601-1613, (1991).
13. Sage, D. *et al.* Quantitative evaluation of software packages for single-molecule localization microscopy. *Nat Methods* **12**, 717-U737, (2015).

The authors would like to acknowledge support by Grant no. 321993 from the Marie Skłodowska-Curie actions of the European Commission, the Lejwa Fund, and Grants no.1417/13 and no. 1937/13 from the Israeli Science Foundation (ES). S.W. acknowledges Willard Chair funds.

# Figure Legends

**Fig. 1. Information build-up of PALM vs. SOFI**

(A) 2nd order SOFI (top row) and PALM (bottom row) images of a representative fixed T-cell expressing Dronpa-Actin and spread on an activating αCD3-coated coverslip. Presented images show the build-up of higher resolution images over the acquisition time. Frame rate: 50 frames/s (20ms exposure time). Bars: 5μm. Bottom left triangles in each image show sum intensity, diffraction limited reconstruction. (B) 2nd order SOFI (top row) and dSTORM (bottom row) images of a representative fixed T-cell, spread on an activating αCD3-coated coverslip, and stained for pZAP70 (αpZAP70 primary Ab) and an Alexa647 secondary Ab. Presented images show the build-up of higher resolution images over the acquisition time. Frame rate: 50 frames/s (10ms exposure time). Bars: 5μm. Bottom left triangles in each image show sum intensity, diffraction limited reconstruction.

**Fig. 2. SOFI-assisted SMLM**

(A) Scatter plots of various PALM localization parameters of localizations within a given pixel vs. SOFI value of the same pixel (plotted for all pixels). Localization parameters include (i) the PSF width ($\sigma$), (ii) the intensity (total number of photons) of the fitted Gaussian, (iii) the local background (BG), (iv) background standard-deviation (STD), and (v) the localization uncertainty ($\hat{\sigma}_{x,y}$). Data was analysed from a movie (same as figure A) of a single representative cell presented in Fig.1A. Manual classification of points was applied to two subpopulations of localizations in the $A_i$ scatter plot (intensity vs. SOFI value) (red and green dots; blue dots represent all localizations). This colouring was next applied to all localizations in all other scatter plots $A_{ii} - A_v$ and for (B) the corresponding PALM image. (C) Scatter plots of various dSTORM localization parameters of localizations within a given pixel vs. SOFI value of the same pixel (plotted for all pixels).

Localization parameters include (i) the PSF width ($\sigma$), (ii) the intensity of the fit Gaussian, (iii) the local background (BG), (iv) background standard-deviation (STD), and (v) the localization uncertainty ($\hat{\sigma}_{x,y}$). Data was analysed from a movie 50 frames/s (10ms exposure time) of a single representative T cell spread on an activating αCD3-coated coverslip and stained with an αpZAP70-Alexa647 antibody. Manual classification of points was applied to two subpopulations of localizations in the $C_i$ scatter plot (intensity vs. SOFI value) (red and green dots; blue dots represent all localizations). This colouring was next applied to all localizations in all other scatter plots $C_{ii} - C_v$ and for (D) the corresponding dSTORM image.

**Fig. 3. Automatic thresholding for background rejection in simulated SMLM data**

(A) The log-normal histogram of SOFI values for 3D simulated data. A threshold was automatically set based on Gaussian fitting of the left-most peak in the histogram. (B) The SOFI image of the simulated data. (C) The SOFI image of the simulated data after automatic thresholding of SOFI values. (D) Analyses of background rejection represented as probability of false detections ($P_{fa}$) vs the detection efficiency represented as the detection probability ($P_d$) for the simulated and automatically filtered data in panels A-C and Fig. S2D-F (SOFI based filtering in blue, sum image based filtering in red).

**Fig. 4. Efficient filtering of SMLM detections using SOFI**

(A) The histograms of SOFI log-values for PALM imaging of a representative T cell expressing Dronpa-actin. A threshold of SOFI value was automatically set based on Gaussian fitting of the left-most peak in the histograms. This threshold formed a spatial mask that excluded all SMLM localizations that were within pixels of SOFI values lower than the threshold, while allowing a 5% background fraction to 'leak' into the image. (B) The SOFI image of the cell. (C) The filtered SOFI image of the cell. (D) The histograms of SOFI log-values for dSTORM imaging of a representative T cell stained for αpZAP70. A threshold was automatically set based on Gaussian fitting of the left-most peak in the histograms, as explained for panel A of this figure. (E) The SOFI image of the cell. (F) The filtered SOFI image of the cell.

**Fig. 5. SOFI-assisted SMLM image reconstruction of live cell imaging**

(A) The probability of detection of a SOFI mask that is dynamically updated during image acquisition of simulated 3D data (shown in Fig. 3B). The SOFI mask was calculated for a sliding window of 100 frames with time steps of 40 frames. The SOFI value threshold for the mask was defined for allowing 5% background fractions (as in Fig. 3A; see further details in the Materials and Methods). (B) Unfiltered (top row) and SOFI-assisted filtered (middle row) PALM images of a representative T-cell expressing TCRζ-Dronpa and spread on an αCD11a-coated coverslip. Presented images show the build-up of higher resolution images over the acquisition time of 50 frames/s (10ms exposure time). Insets show zoomed regions where out-of-focus background broadens the cell features in the unfiltered images. Intensity profiles of features in the insets (black angled lines in the insets) are shown in the bottom row. Bars – 5μm (full image) and 0.5μm (insets). Colour bars for top row images (left to right) – 0 to 16592, 4463, or 30984 localizations/μm$^2$. Colour bars for middle row images (left to right) – 0 to 5554, 3311, or 3485 localizations/μm$^2$. Each image in the time series is the sum of 100 single frames analysed by ThunderSTORM and SOFI filtered as described above for live cells.

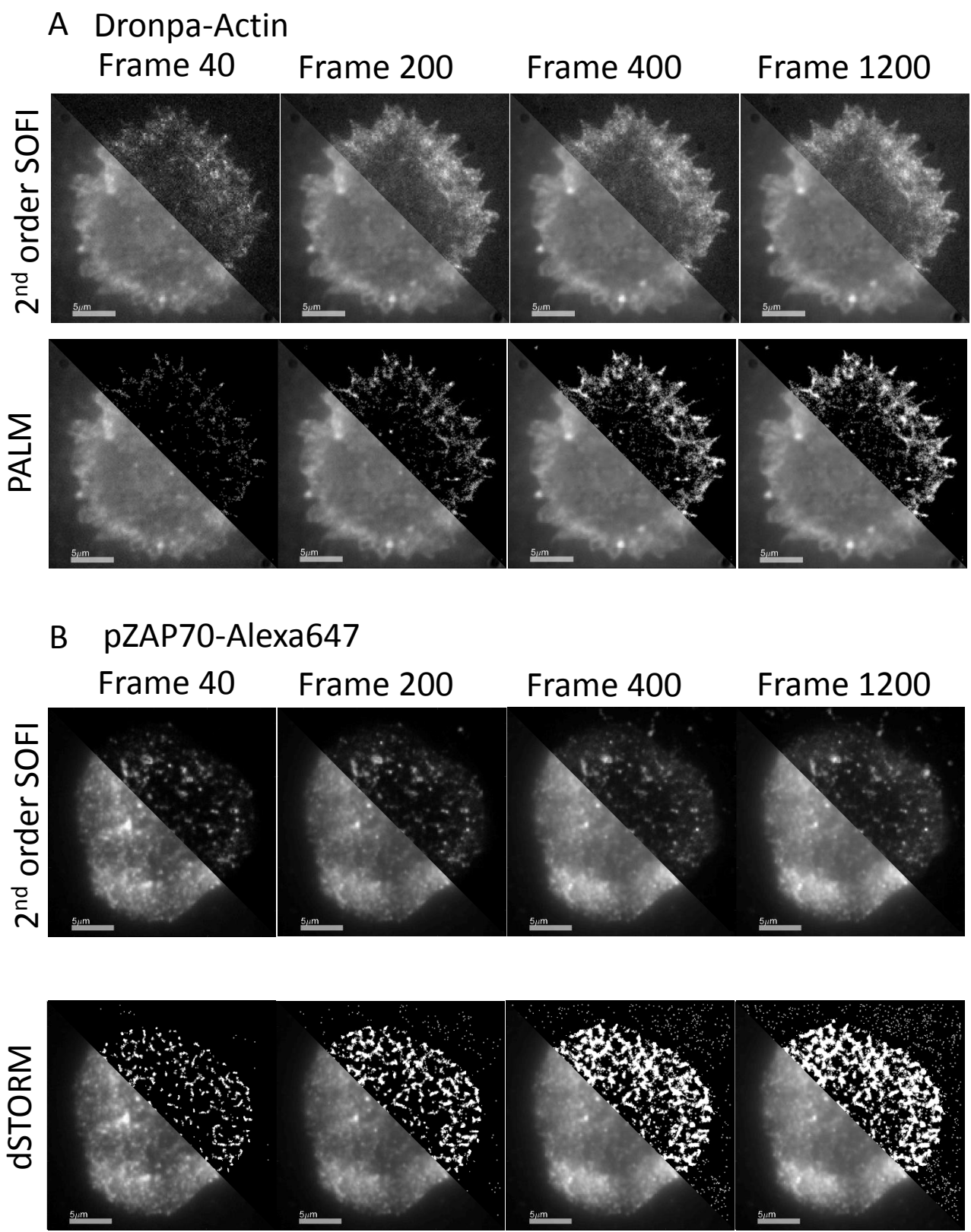

Fig. 1. Schidorsky, Yi, et al.

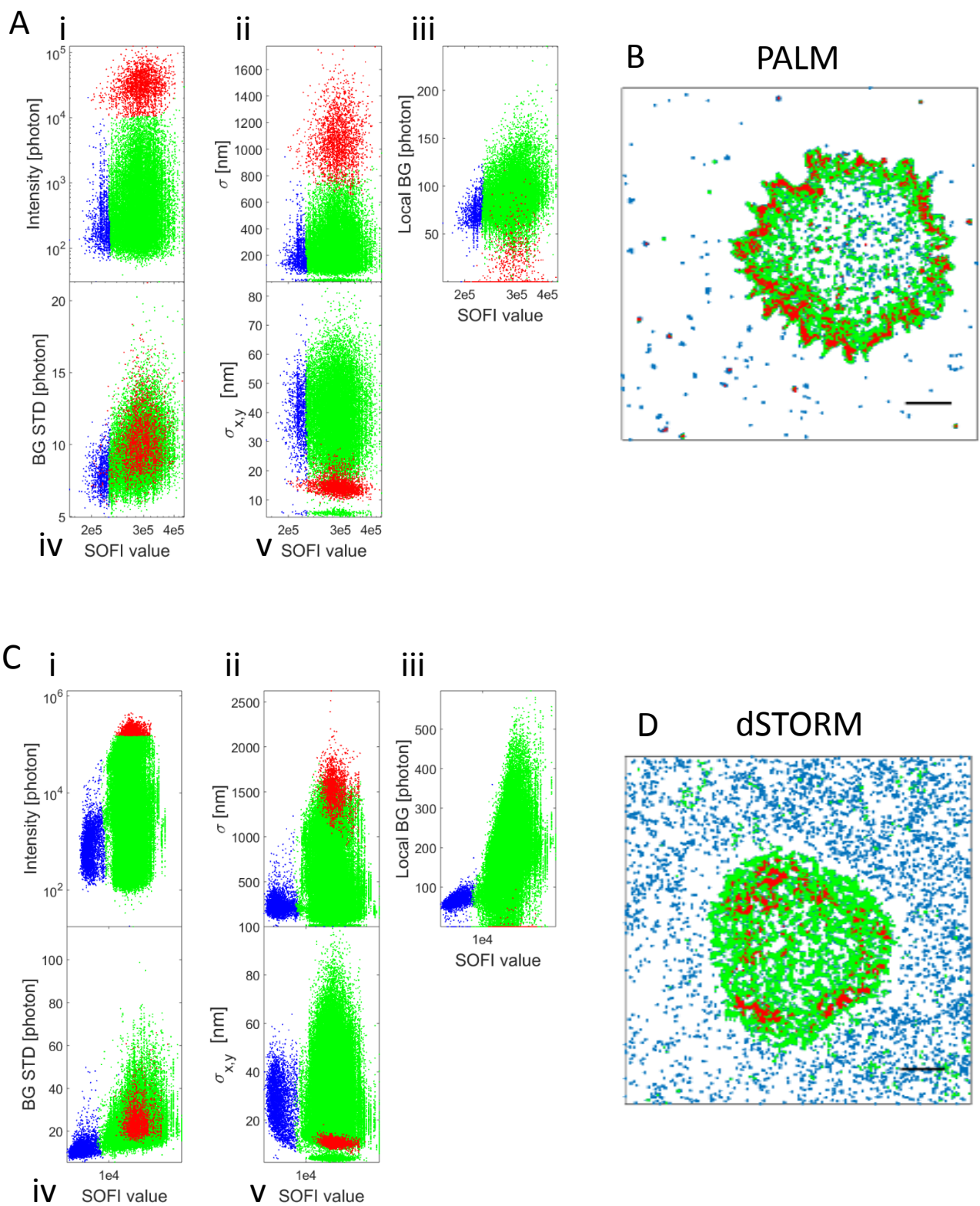

Fig. 2. Schidorsky, Yi, et al.

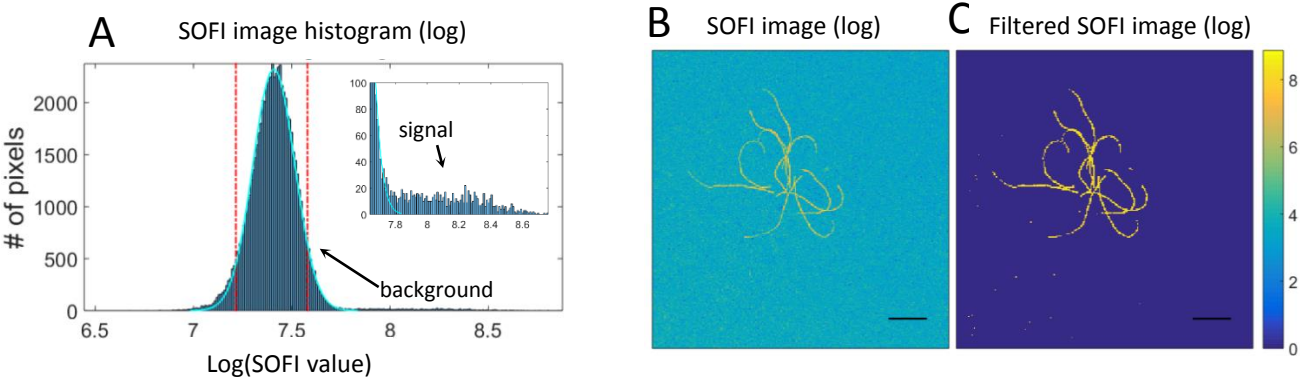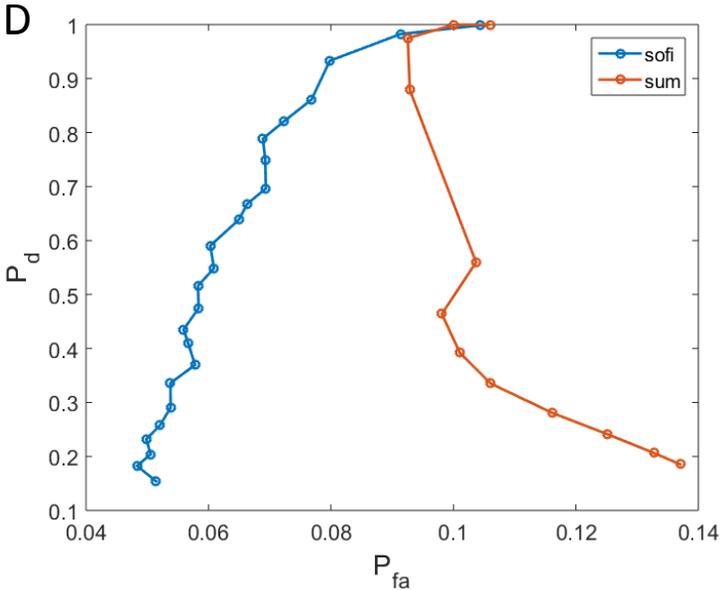

Fig. 3.    Schidorsky, Yi, et al.

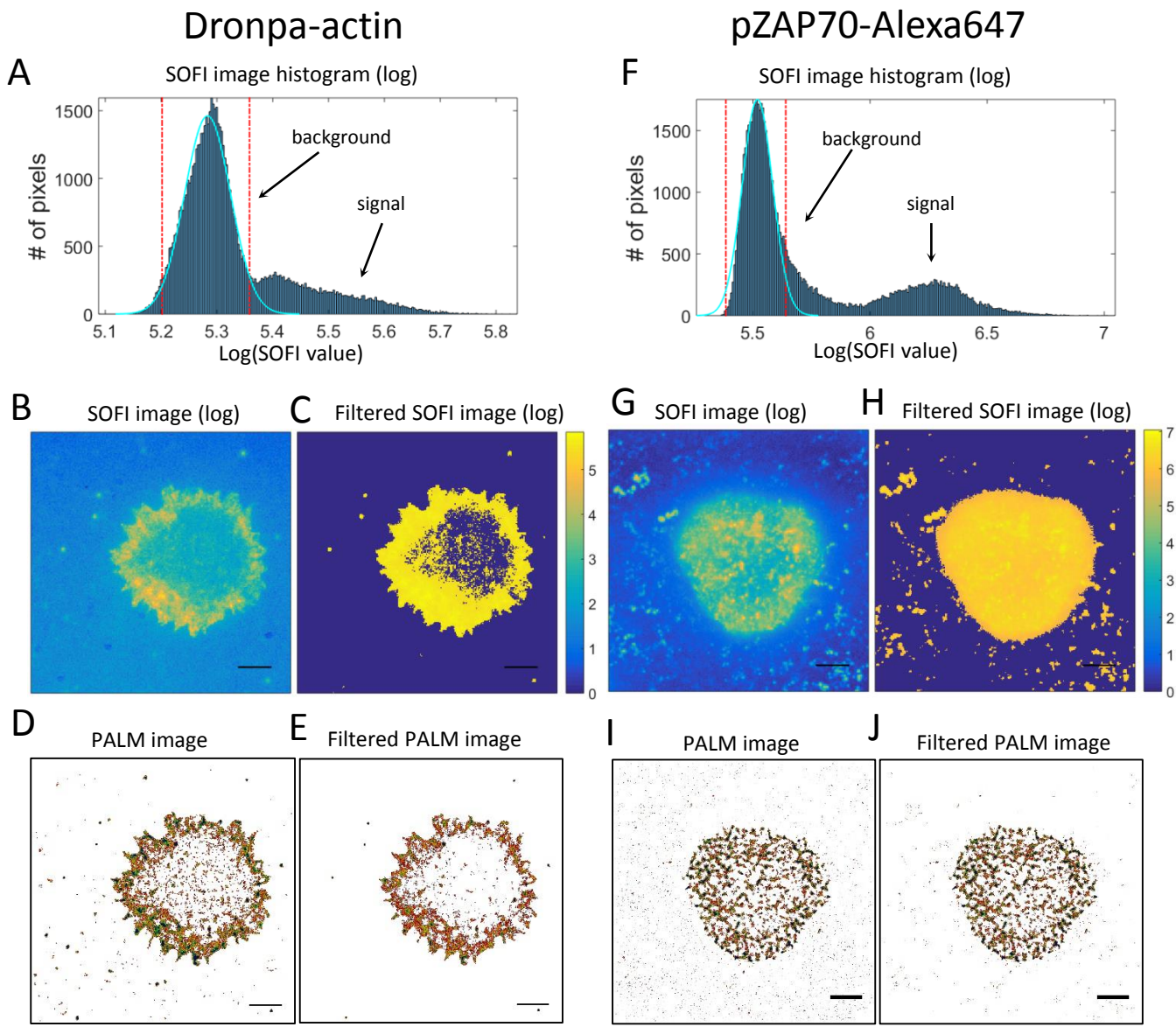

Fig. 4. Schidorsky, Yi, et al.

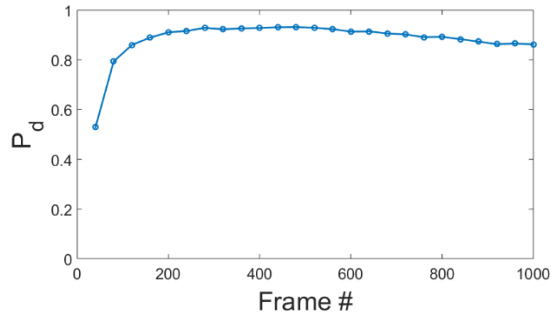
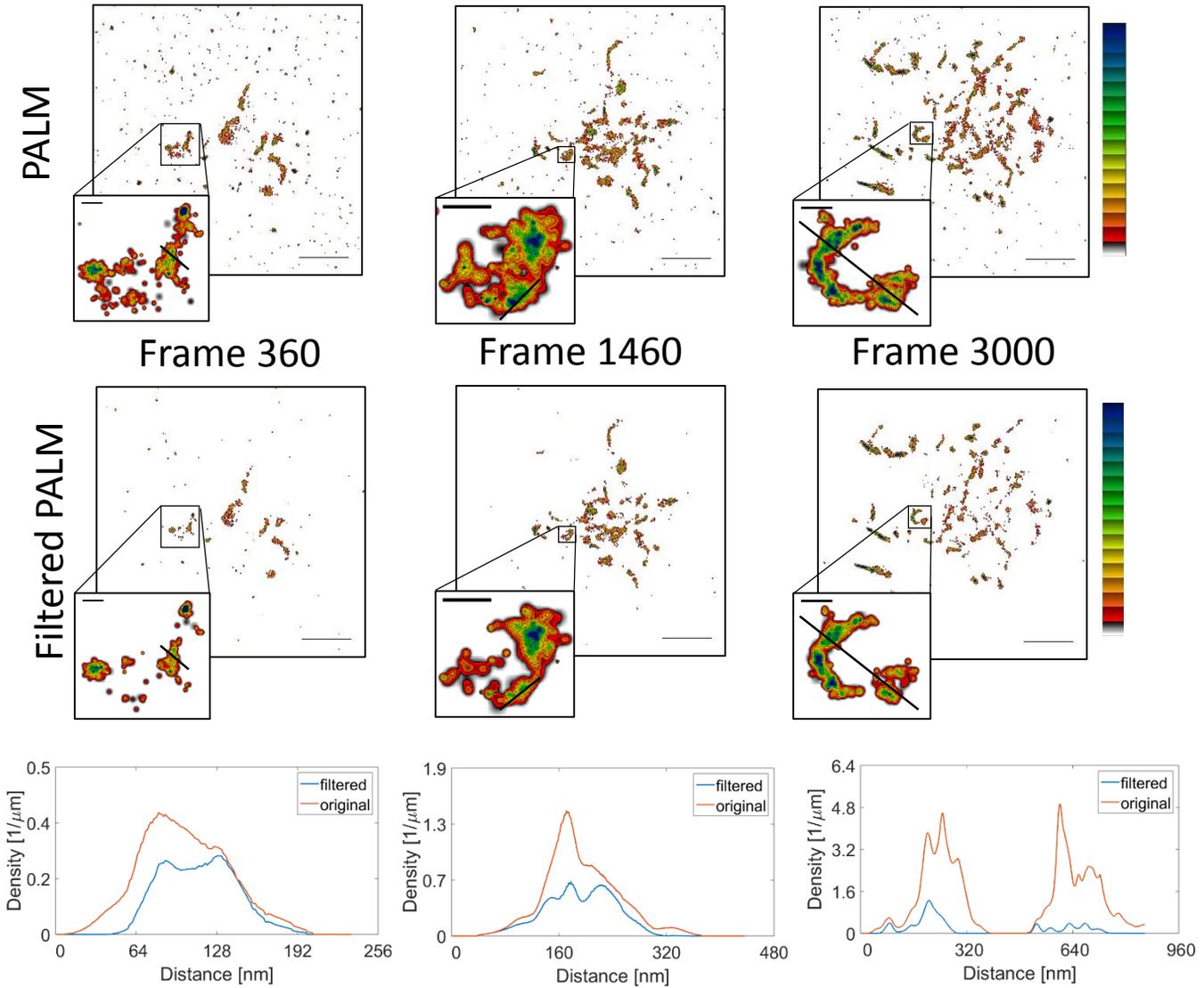

Fig. 5. Schidorsky, Yi, et al.

A    2D simulation          B    3D simulation

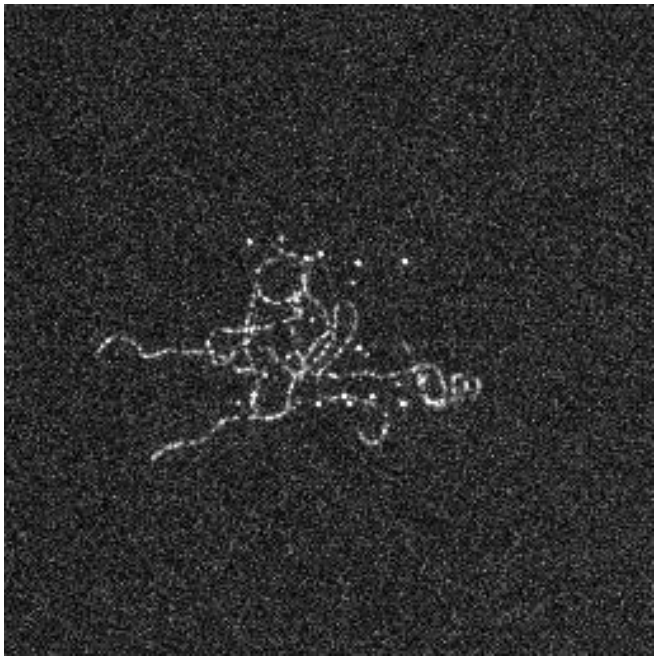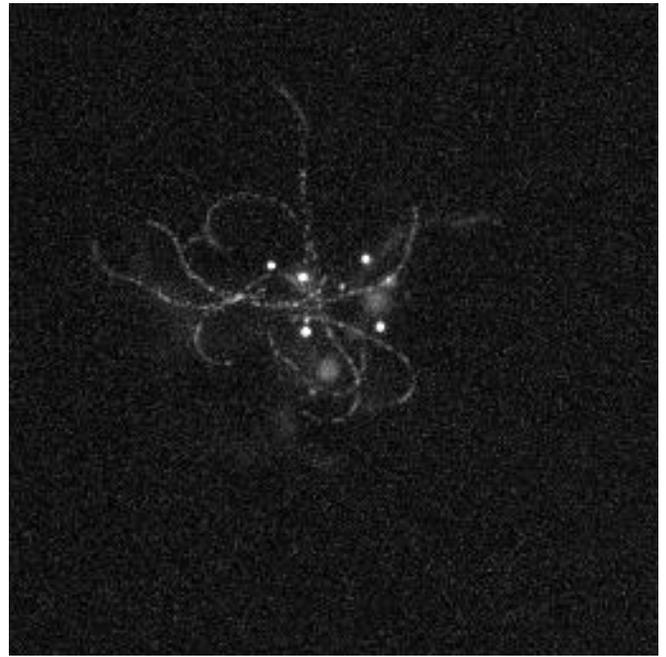

C
i   ii   iii

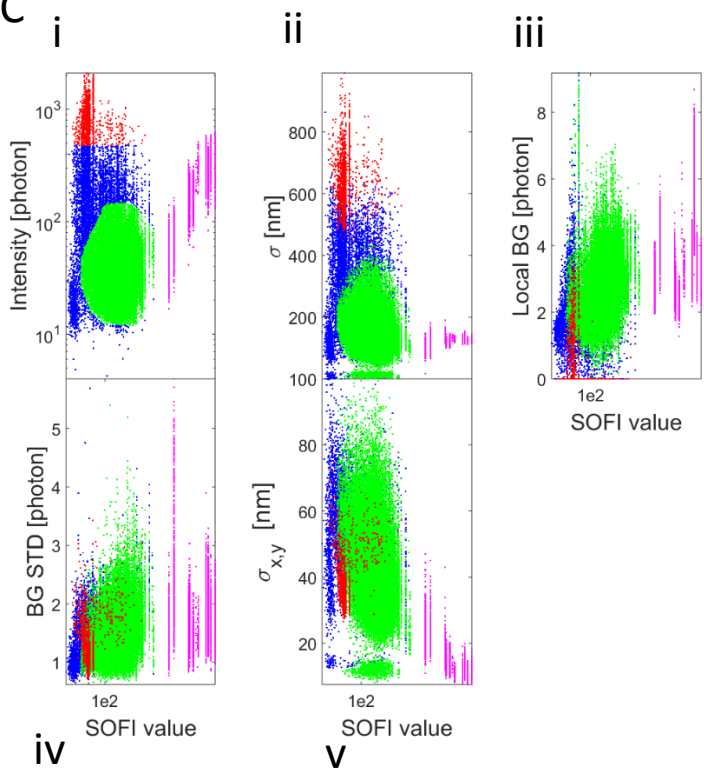

iv   v

D    3D simulation

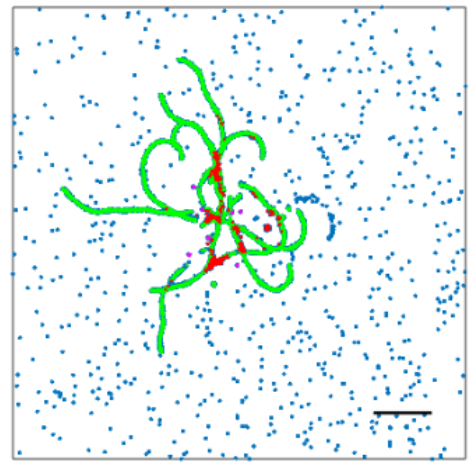

Fig. S1.    Schidorsky, Yi, et al.

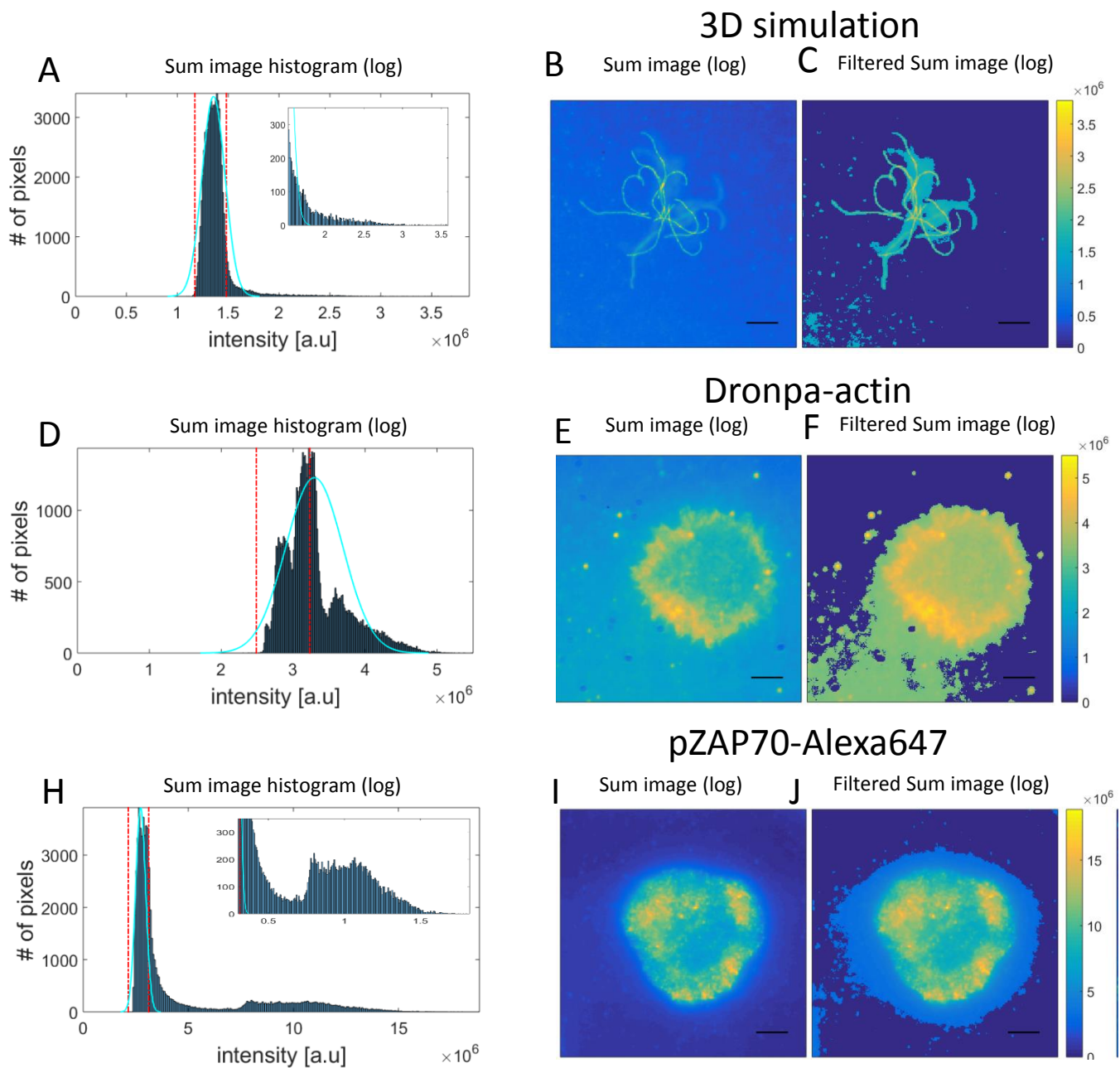

Fig. S2. Schidorsky, Yi, et al.

# Supplementary Information for:

# Synergizing superresolution optical fluctuation imaging with single molecule localization microscopy

Shachar Schidorsky[1,3], Xiyu Yi[2,3], Yair Razvag[1], Yonatan Golan[1], Shimon Weiss[2]*, Eilon Sherman[1]*

[1]*Racah Institute of Physics, The Hebrew University, Jerusalem, Israel, 91904*

[2] *Department of Chemistry & Biochemistry, UCLA, CA, USA, 90095-1569*

[3] *Equal contribution*

*\* Corresponding authors*

**Outline**

    **1. Materials and Methods**

    **2. Realistic simulation – General description**

    **3. Supplemental Figure Legends**

    **4. References**

# 1. Materials and Methods

**Cloning**

Actin tagged with the photoactivatable fluorescent protein (PAFPs) Dronpa[1] (MBL International Corporation) was cloned in EGFP-N1 or EGFP-C1 vectors (Clontech). PAFPs genes were cut from a vector through digestion with restriction enzymes (typically AgeI and NotI, XbaI or BsrGI). PAFP genes then served to replace existing fluorescent proteins (FP) in previously used constructs (1-2) using similar digestion reactions and ligation of the PAFP insert (Quick-ligation kit, New England BioLabs). Validation of cloning was done by restriction digestion analyses and DNA sequencing of the inserts.

**Cell line**

E6.1 Jurkat T cells were transfected with DNA using a Neon Electroporator. Transiently transfected cells were maintained in transfection medium, sorted for positive expression of Dronpa chimeras and imaged within 48-72 hours from transfection. Sorting was done at the flow cytometry center of the Silberman Institute on a high speed cell sorter (Becton Dickinson flow cytometer). Stable cell lines were created by selection with Geneticin at 1.5mg/ml (G418, Invitrogen). After 2-3 weeks, the cells were sorted for positive protein expression by flow cytometry. Cells were then evaluated using biochemistry assays, flow cytometry, and light microscopy (epifluorescence, TIRF and PALM imaging, as described below).

**Sample preparation**

The preparation of coverslips for imaging spread cells followed a previously described technique (e.g. [2]). Briefly, for diffraction limited and PALM imaging, coverslips (#1.5 glass chambers, LabTek or iBidi) were washed with acidic ethanol at room temperature (RT) for 10 min; liquid was then aspirated and coverslips were dried at 40°C for 1 hour. Cleaned coverslips were incubated at RT for 15 min with 0.01% poly-L-lysine (Sigma) diluted in water. Liquid was aspirated and coverslips were dried at 40°C for 12 hours. Coverslips were subsequently incubated with stimulatory or non-stimulatory antibodies at a concentration of 10 mg/ml (unless specified otherwise) overnight at 4°C or 2 hours at 37°C. Finally, coverslips were washed with PBS. Throughout the study we used the following stimulatory antibodies: purified mouse anti human αCD3 (clone Ucht1) and αCD11a (BD Biosciences). A few minutes before imaging, cells were resuspended in imaging buffer at a concentration of 1 million/150 ml and 100,000-500,000 cells were dropped onto coverslips for PALM or diffraction limited imaging, incubated at 37°C for the specific spreading time (typically 3 min) and fixed with 2.4% PFA for 30 min at 37°C or used for live cell imaging.

**PALM microscopy**

Photoactivated localization microscopy was conducted similarly to the imaging previously described (e.g. [3]), using a total internal reflection (TIRF) Nikon microscope (Eclipse Ti). Dronpa-tagged proteins were activated using a low intensity laser illumination at 405 nm (~1%), and sequentially imaged in a following frame using laser excitation at 488 nm. Both laser illuminations were in TIRF mode to minimize background from the cell cytosol. Dronpa-tagged molecules were imaged for up to 3 min for live-cell imaging, or the depletion of their emission, as identified by the loss of fluorescence, for fixed-cell imaging. The focus of the microscope was maintained throughout the imaging using the PerfectFocus system of the microscope.

**Immunostaining for dSTORM**

Zap70 proteins were labelled using dye-labeled goat anti-rabbitt secondary antibodies. Antibodies were labelled using the manufacturers' protocols. In brief, 0.5 µg of ZAP-70 [pY315/pY319] monoclonal antibody (Life Technologies) was added to 500×103 cells suspended in FACS buffer for 30 min on ice. Then cells were washed two times in PBS and suspended in 1.5 ml of FACS buffer containing 1 µg of goat anti-rabbit IgG (H+L) secondary antibody, Alexa fluor 647 conjugate (Life Technologies). Cells were washed and suspended in imaging buffer.

**dSTORM microscopy**

For direct stochastical optical reconstruction microscopy of single-molecule fluorophores cells were fixed and suspended in a STORM imaging buffer which was made by a protocol previously described [4]). Imaging was performed using a total internal reflection (TIRF) Nikon microscope. Alexa 647-tagged proteins were activated using a low intensity laser illumination at 405 nm (~0.5%), and sequentially imaged in a following frame using laser excitation at 647nm. Both laser illuminations were in TIRF mode to minimize background from the cell cytosol. Alexa 647-tagged molecules were imaged for up to depletion of their emission, as identified by the loss of fluorescence. The focus of the microscope was maintained throughout the imaging using the PerfectFocus system of the microscope.

**Data analyses**

Generation of super-resolved images

Movies generated by SMLM acquisitions (either PALM or dSTORM imaging) were analyzed by the ThunderStorm software [5] for the identification of individual molecules in the movie frames and for rendering of the super-resolved images. SMLM peak detection by ThunderSTORM involved the following steps. SMLM image filtering was performed using a wavelet filter (B-SPLINE order 3). Localization was performed by identifying local maxima with 8-neighbourhood connectivity. Sub-pixel localization further employed the integrated Gaussian in the PSF model with fitting radius of 3 pixels, the weighted least-square fitting option, and initial sigma of 1.4 pixels. Individual molecules were rendered with intensities that correspond to their fitted Gaussian with respect to the maximal probability density values detected in the field. Image rendering utilized an overlay of normalized Gaussians.

To generate SOFI images from the acquired data, $2^{nd}$-order SOFI analysis with time lags equal to zero was used throughout the study. SOFI analyses and image reconstruction were performed using the 'Localizer' software [6].

Automated threshold detection

The automatic threshold was determined by the following algorithm:
1. Perform SOFI analysis of the wanted frames in the movie – in our case we used $2^{nd}$ order A.C SOFI.

2. Compute the histogram of the log of the SOFI image using the function "histogram" in MATLAB.
3. Find the first peak using the function findpeaks with "MinPeakProminence" parameter set to 100.
4. Use the width output parameter from findpeaks to determine borders in histogram for Gaussian fitting.
5. Fit Gaussian.
6. Use norminv MATLAB function to calculate how much of the fitted Gaussian (that is the assumed background) to include and how much to cut off. For example, for a value of $Cutoff = 0.05$, we can detect the threshold by: $norminv(1 - Cutoff, \mu, \sigma)$. Where $\mu$ and $\sigma$ are obtained from the fitted Gaussian and the cutoff is determined by the researcher.
7. Apply the cutoff to the SOFI image and create a bitmask.
8. Apply the bitmask to the PALM image

In our study, we applied the same steps (1-8) to generate masks based on sum intensity images. This allowed the comparison of our SOFI-assisted masking to a potentially simpler approach of SMLM masking based on sum-intensity images (Fig. S2). This latter approach showed consistently a poor background-rejection capability, in comparison to SOFI-based masking.

We note that SOFI inherently assigns low values to uncorrelated background, relative to the more self-correlated signal that scores higher values. Under the observation that the lowest SOFI-values constitute the background, our algorithm excludes efficiently true-negative detections, while allowing a predetermined fraction of the background (false-positive detections) to leak through the mask.

Evaluation of mask performance

Following detection theory, we can evaluate the performance of the SOFI masking of SMLM data using the following parameters of detection (used here interchangeably with SMLM localizations). One useful parameter is the probability of detections ($P_d$) and the other is the probability of false detections (i.e. false alarms, $P_{fa}$). Note that true negative detections, due to correctly rejected background, are of little interest in our case. Since our goal is to evaluate the performance of the mask in relation with SMLM detections, we turn to simulation where the ground-truth emitters (GT) are well defined. Then, we employ ThunderSTORM to localize the single emitters from the simulated data. We can now define the fraction of the ground-truth detected emitters (GT_D). Next, we employed SOFI-assisted masking to the SMLM localizations (found by ThunderSTORM) and define the fraction of GT_D that passed the mask (Filtered_GT_D). Now, we can define the detection parameters in the following way:

$$P_d = \frac{\text{Filtered\_GT\_D}}{\text{GT\_D}}$$

The SMLM localizations contain also false detections. Hence, we need to define the whole set of detections that went through the SOFI mask (Filtered detected emitters, Filtered_D). Now the false detections are (Filtered_D – Filtered_GT_D), and the $P_{fa}$ can be defined as:

$$P_{fa} = \frac{\text{Filtered\_D} - \text{Filtered\_GT\_D}}{\text{Filtered\_D}}$$

Similar parameters can be defined for an arbitrary binary mask. A specific mask of interest is based on sum-intensity thresholding, as described earlier in this section and in Fig. S2.

SOFI imaging

Following Dertinger, et al. 2009 [7], we define 2nd order SOFI as follows. Assuming the PSF of our optical system is denoted by U(r), the intensity of a pixel in time t, and location r, can be described by the following equation: $F(\vec{r}, t) = \sum_{k=1}^{N} U(\vec{r} - \vec{r_k}) \cdot \epsilon_k \cdot s_k(t)$, where k denotes the index of a single emitter, N is the total number of emitters, $\vec{r_k}$ is the location of emitter k, $\epsilon_k$ is the intensity of emitter k and $s_k(t)$ is a binary function that approximates the blinking dynamics of emitter k. Next, we can define the intensity fluctuation as $\delta F(\vec{r}, t) = F(\vec{r}, t) - \langle F(\vec{r}, t) \rangle_t = \sum_k U(\vec{r} - \vec{r_k}) \cdot \epsilon_k \cdot \delta s_k(t)$. Given these definitions, 2nd order SOFI is defined as the 2nd order auto-correlation of the intensity fluctuations:

$$AC_2 \equiv \langle \delta F(\vec{r}, t+\tau) \cdot \delta F(\vec{r}, t) \rangle_t = \sum_j U^2(\vec{r} - \vec{r_k}) \cdot \epsilon_k^2 \cdot \langle \delta s_k(t+\tau) \delta s_k(t) \rangle_t .$$

Calculating this equation for every pixel in the image with $\tau = 0$, we get the "SOFI image". Assuming the PSF can be effectively approximated by a Gaussian, i.e. $U(\vec{r}) = \exp\left(-\frac{x^2+y^2}{2\omega_0^2}\right)$, we get an increase of $\sqrt{2}$ in the effective resolution of the image as well as the effective elimination of non-correlated pixel intensities over time. Effective elimination means that non-correlated intensity fluctuations will get very low correlation values in the SOFI image. Thus, we observe that the first peak in the histogram of SOFI pixel values correspond to the background pixels (e.g., see Fig. 4 A,F).

Gray level conversion for SMLM rendering

SMLM images are rendered in gray levels (e.g. Fig. 1A) or in false colors (Fig.5B). The gray levels represent the probability density of finding an emitter in a pixel. Thus, the gray level values should be considered with units of $\frac{localization}{\mu m^2}$. To illustrate objects in SMLM images, we used a conversion of gray levels to false coloring using '16 ramp' look-up table in ImageJ.

## 2. Realistic simulations

In order to have a good ground-truth data to help with the development, evaluation and optimization of our propose method, we implemented a realistic simulator that generates simulations of fluorescence microscopy data and incorporates the realistic considerations of sample conditions, photo-physical properties, and imaging system conditions of an actual experimental dataset. The entire simulator is home-built using MATLAB (R2012a). A script version with separate modules as MATLAB functions are available in SI software package. Below, we describe the "bottom-up" construction of the simulator, based on six major steps in a pipeline.

**2.1. Modeling feature of interest: random filaments.**

As the first step in our simulation, we lay down the feature of interest with continuous 3D coordinates that mimic randomly curved filaments, with a given virtual sample volume. These filaments structure supposed to carry certain labeling sites of interest. It can represent actin or tubulin filaments within mammalian cells, filaments in the immune synapse, or other features.

In the actual algorithm, one random curve is generated as a simulated chain described in the following steps (MATLAB script available in SI software package):

First, based on the user-defined curve length and available labeling site density that is feature dependent, we calculate the total number of available labeling site (denoted as 'nodes' in the following discussion) and the distance between adjacent nodes (denote as 'node distance' in the following discussion) on the curve. Then, starting from the first node located at origin, we put down node locations one after another until we have reached the total number of nodes. Starting from the first node, the location of the next node is defined by the location of the previous node, node distance and the elongation direction. The elongation direction is defined by two factors: First, the elongation direction in the previous node; second, a further tilt angle that in a 3D polar coordinate system that incorporates short and long range changes of elongation angles. $\theta_n$ and $\varphi_n$ are the two angles that defines the elongation direction after $n^{th}$ node in 3D polar coordinate. $\Delta L$ is the distance between two adjacent nodes. $x_n$, $y_n$, $z_n$ are the 3D coordinates of the location of the $n^{th}$ node. N is the total number of nodes. $P_1$, $P_2$, $P_3$, $P_4$ are free tunable parameters that can be selected at wish. $C\theta_n$ and $C\varphi_n$ (weighted by parameter $P_1$ and $P_3$) are cumulative angle changes, that gives rise to a long range bending feature that has increasing curvature for each simulated filament, and the actual tilt angle after each node is further perturbed by another small angle weighted by parameters $P_2$ and $P_4$. One simulated filament is thus computed in an iterative manner that construct one node after another, until the total number of nodes is reached.

Repeating the process listed above can generate multiple filaments, thus we can control the virtual sample to have either sparse or dense feature of interest. In order to save on computational cost one can also generate a library of sparse features, and overlay multiple features from different simulations to get a dense feature (one such library is available for downloading at Github, www.github.com/xiyuyi/RealisticSimulator_3D). The output of this step in the simulation is a list of 3D coordinates encoding the locations of available labeling sites.

**2.2 Modeling emitters' distribution on (and off) features: Labeling density, Labeling uncertainty, non-specific background and Aggregates.**

Once features of interest are placed inside the virtual sample volume, the next step constitute the placement of emitters ('labeling') on features of interest. Experimentally, labeling is usually performed by either immunostaining with primary/secondary antibodies, or by fusion of fluorescence proteins (FPs) to target protein(s) of interest. Labeling density on features of interest is often limited by copy number, epitope availability, antibody size etc. For example the diameter of flexible filaments is approximately 7nm [8]. Labeling density of T-cell receptors is expected to be in the range of 150 $\frac{molecule}{\mu m^2}$ [3]. The simulator allows tuning these parameters. As the physical length of each filament is determined in the previous step, tuning of labeling density is achieved via a choice of 'labeling ratio' $\equiv \frac{\#\ of\ occupied\ labeling\ sites}{total\ number\ of\ available\ labeling\ sites}$. The sites to be occupied are picked up randomly. For filaments, labeling density translates into number emitters per unit length. These chosen sites are the selected labeling sites. Before the actual placing of emitters on selected labeling sites, we also need to account for antibody (fusion protein) size, linker length, orientation of epitopes etc. These parameters give rise to a distribution of emitters' distances around the filament. We therefore introduce an additional random parameter, dubbed "Labeling uncertainty $\sigma_u$". The coordinates corresponding to each selected labeling site are modified by a vector $\Delta \vec{r}$ that follows Gaussian distribution $P(\Delta r) = \exp\left(-\frac{\Delta \vec{r}}{2\sigma_u^2}\right)$. Nonspecific labeling of emitters (a common problem in immunolabeling) is simulated by the addition of a chosen number (volumetric density) of randomly picked 3D coordinates (off the filaments). Aggregates of emitters (which are often encounters with FPs) were simulated by picking up a high local density of emitters within a confined 3D volume, with a chosen aggregate size and a chosen emitter density within the aggregate. This module outputs a list of 3D coordinates that indicate the location of all the emitters in the virtual sample volume. The next module simulates fluctuating ('blinking') time-intensity trajectory for each of the emitters in the list.

**2.3 Modeling realistic blinking statistics and bleaching effects.**

For each single emitter, the blinking profile is constructed based on emission on/off of Dronpa, as reported by S. Habuchi [9]. Specifically, typical fluorescence of a single-molecule of Dronpa was observed to exhibit individual intense bursts, separated by long off-times in the order of tens of seconds. Within the bursts, blinking statistics was characterized by medium off-times, short off-times and short on-times. Thus our simulation is based on six time constants, that characterize the averages on-time distribution ($\tau_{on}^*$), long-off time distribution ($\tau_{l-off}^*$), medium-off time distribution ($\tau_{m-off}^*$), short-off time distribution ($\tau_{s-off}^*$), the standard deviation of short-off time distribution ($\sigma_{s-off}^*$) as well as the average duration of bursts between long-off times for a trajectory ($\tau_{bst}^*$). Short-off time series was calculated with Gaussian distribution with average $S_{s-off}^*$. For on time, long-off time, and medium-off time series that follows exponential distribution, we calculated the series as $\tau = -\frac{1}{\tau^*}\ln(1-x)$, where x follows a uniform distribution in the range of [0, 1], and $\tau^*$ represents $\tau_{on}^*$, or $\tau_{m-off}^*$ or $\tau_{l-off}^*$. Each blinking trajectory (without bleaching) was constructed by first calculating long-off time series such that the total time of the series covers the time range of a simulation movie, because $\tau_{l-off}^*$ is typically longer than $\tau_{m-off}^*$, $\tau_{s-off}^*$ and $\tau_{on}^*$. Between each long-off time we inserted trajectories of bursts, 35% of which are constructed with both on-times and long-off times, and the rest (65%) with on-times and short-off times. Duration of each burst was restricted to $\tau_{bst}$ that follows exponential distribution at defined average $\tau_{bst}^*$. For each burst, the initial and ending state is fixed to be the on state. If the initial on state is longer than the burst duration then the burst is kept with only one on-time period. If the summation of initial on-time and off-time is greater than the burst duration, then the next on-time period is kept. Such extension of bursts durations slightly alters the bursts duration distribution, but maintains the distribution of on/off times. Intensity of the trajectory in a single frame is calculated as the ratio of on time of the frame with integration time tint, multiplied by given SNR.

Bleaching effects are simulated by randomly turning active emitters into "bleached" emitters after which the emitter will no longer be active. The selection of emitters was set to maintain the percentage of active emitters a(t) such that a(t) is proportional to B(t), where B(t) is the experimentally measured bleaching profile, or pre-computed bleaching profile. This module outputs a list of time-intensity trajectories corresponds to every single emitter in the simulation volume.

**2.4 Realistic 3D Modeling: PSF calculation, out-of-focus light, and emitter brightness distribution.**

The outputs of the modules described above provide 3D coordinates for each emitter and a blinking time-intensity trajectory for each emitter. The brightness of the emitters could be set as identical, randomly picked up from a preset brightness distribution, or follow an exponential decay pattern that mimics TIRF excitation. The system-response of the optical setup is accounted for by propagating the emitted signals (point sources) from the sample plane(s) to the detector plane. We first calculate the 3D PSF function based on Gibson Lanni's PSF model using an open source software (PSF Generator, available at 'http://bigwww.epfl.ch/algorithms/psfgenerator') with specified optical setup parameters. Then a focal plane 'slice' of depth z is chosen in the sample volume. For every single emitter, the difference between its z-axis coordinates with the z-coordinate of the focal plane is calculated and used to choose it's corresponding 2D PSF (from the 3D-PSF stack). If the emitter is not located in the focal plane slice, its 2D-PSF will be catalogued as an 'out-of-focus' PSF. The program goes through every single emitter, multiplies the pre-computed brightness of each emitter with its corresponding 2D-PSF, and additively transfers the image of each emitter to the imaging matrix, to form a diffraction limited image on the detector with a 10 x 10 finer pixelated grid (as compared to sample plane's grid), representing the camera. Noting that the emitter location was simulated in a 3D volume, so the emitters have 3D coordinates, that will give every emitter a different 2D-PSF, based on the corresponding 3D section on the simulated 3D-PSF. Under this manner, the out-of-focus light is naturally encoded in the simulation. More Out-of-focus emitters can be added by selecting more random locations in the simulated 3D volume and follow the same image construction scheme described above.

**2.5 Modeling binning effects, noise, and background signals.**

The 10x10 binning is applied to each movie frame to simulate the camera binning effects. A Poisson noise is then simulated by re-setting each pixel by a random value from a Poisson distribution having a mean value which is equal to the noise-free pixel value.

**2.6 Addition of Measured Background to Simulation**

After the preparation of the simulated signal, background was added by using recorded frames from a sample that did not include cells. The sample was prepared using an identical protocol to that for fixed cells (see sample preparation section above) without adding cells. Before adding the background, we could change the signal-to-background (SBR) ratio by multiplying the signal with a desired factor. The performance of our SOFI-assisted background rejection in SMLM imaging was evaluated under various SBR conditions (0.2-5). We present results (Fig. 3) for an SBR of 0.6, which constituted relatively difficult conditions for SMLM detection.

## 3. Supplemental Figure Legends

**Fig. S1. Realistic simulations of SOFI / SMLM movies**

(A) A representative simulated SMLM frame (from a movie) of tubulin-like 2D structures. Simulations included Gaussian noise, protein aggregates, and TIRF excitation. (B) A representative simulated SMLM frame (from a movie) of tubulin-like 3D structures. Simulations included Gaussian noise, protein aggregates, out of focus light, and epi-illumination excitation. (C) Scatter plots of various SMLM localization parameters of localizations within a given pixel vs. SOFI value of the same pixel (plotted for all pixels). Localization parameters include (i) the PSF width ($\sigma$), (ii) the intensity of the fit Gaussian, (iii) the local background (BG), (iv) background standard-deviation (STD), and (v) the localization uncertainty ($\hat{\sigma}_{x,y}$). Data was analysed from a movie (same as figure A) of simulated data presented in Fig.S1D. Manual classification of points was applied to three subpopulations of localizations in the $A_i$ scatter plot (intensity vs. SOFI value) (red, green and magenta dots; blue dots represent all localizations). This colouring was next applied to all localization in all other scatter plots $C_{ii} - C_v$ and for (D) the corresponding SMLM image.

**Fig. S2. Inefficient filtering of SMLM localizations using sum-image thresholding.**

(A) The histogram of sum-image log-values for 3D simulated data. A threshold was automatically set based on Gaussian fitting of the left-most peak in the histogram. (B) The sum-image of the simulated data. (C) The sum-image of the simulated data after automatic thresholding of sum-image values (see details on automatic thresholding in section on data analyses in the SI). (D) The histograms of sum-image log-values for PALM imaging of a representative T cell expressing Dronpa-actin. A threshold was automatically set based on Gaussian fitting of the left-most peak in the histograms. (E) The sum-image of the cell. (F) The filtered sum-image of the cell. (G) The histograms of sum-image log-values for dSTORM imaging of a representative T cell stained for

αpZAP70. A threshold was automatically set based on Gaussian fitting of the left-most peak in the histograms. (H) The sum-image of the cell. (I) The filtered sum-image of the cell.

**Movie M1. The dynamics of live cell PALM imaging with and without SOFI-assisted background filtering**

PALM imaging of a representative T cell expressing TCRζ-Dronpa with (right) and without (left) SOFI-assisted filtering. A threshold was automatically set based on Gaussian fitting of the left-most peak in the histograms. Each single frame in the attached movie is the summation of all the detected peaks within a moving window of 220 frames centred at each frame). The film was saved using ImageJ. Scale bar - 5μm. False-colouring utilized a "16 ramps" lookup-table (LUT; obtained from http://imagej.nih.gov/ij/download/luts/).

## 4. References

1   Ando, R., Mizuno, H. & Miyawaki, A. Regulated fast nucleocytoplasmic shuttling observed by reversible protein highlighting. *Science* **306**, 1370-1373, (2004).
2   Sherman, E. *et al.* Functional nanoscale organization of signaling molecules downstream of the T cell antigen receptor. *Immunity* **35**, 705-720, (2011).
3   Neve-Oz, Y., Razvag, Y., Sajman, J. & Sherman, E. Mechanisms of localized activation of the T cell antigen receptor inside clusters. *Biochim Biophys Acta* **1853**, 810-821, (2015).
4   Dempsey, G. T., Vaughan, J. C., Chen, K. H., Bates, M. & Zhuang, X. Evaluation of fluorophores for optimal performance in localization-based super-resolution imaging. *Nat Methods* **8**, 1027-1036, (2011).
5   Ovesny, M., Krizek, P., Borkovec, J., Svindrych, Z. & Hagen, G. M. ThunderSTORM: a comprehensive ImageJ plug-in for PALM and STORM data analysis and super-resolution imaging. *Bioinformatics* **30**, 2389-2390, (2014).
6   Dedecker, P., Duwe, S., Neely, R. K. & Zhang, J. Localizer: fast, accurate, open-source, and modular software package for superresolution microscopy. *Journal of biomedical optics* **17**, 126008, (2012).

7	Dertinger, T., Colyer, R., Iyer, G., Weiss, S. & Enderlein, J. Fast, background-free, 3D super-resolution optical fluctuation imaging (SOFI). *Proc Natl Acad Sci U S A* **106**, 22287-22292, (2009).
8	Cooper, G. M. *The Cell - A Molecular Approach 2nd Edition*.  (Sunderland (MA): Sinauer Associates, 2000).
9	Habuchi, S. *et al.* Reversible single-molecule photoswitching in the GFP-like fluorescent protein Dronpa. *Proceedings of the National Academy of Sciences of the United States of America* **102**, 9511-9516, (2005).